\documentclass[prl,twocolumn,superscriptaddress,showpacs]{revtex4}
\usepackage{amsmath}
\usepackage{amsfonts}

\begin{document}
\title{Entanglement constrained by superselection rules} 
\author{Stephen D. Bartlett} 
\email{bartlett@physics.uq.edu.au} 
\affiliation{Department of Physics, Macquarie University, Sydney, New
  South Wales 2109, Australia}
\author{H. M. Wiseman} 
\email{H.Wiseman@griffith.edu.au}
\affiliation{Centre for Quantum Computer Technology, Centre for
  Quantum Dynamics, School of Science, Griffith
  University, Brisbane, Queensland 4111, Australia}
\date{26 August 2003}

\begin{abstract}
  Bipartite entanglement may be reduced if there are restrictions on
  allowed local operations. We introduce the concept of a generalized
  superselection rule (SSR) to describe such restrictions, and
  quantify the entanglement constrained by it.  We show that ensemble
  quantum information processing, where elements in the ensemble are
  not individually addressable, is subject to the SSR associated with
  the symmetric group (the group of permutations of elements). We
  prove that even for an ensemble comprising many pairs of qubits,
  each pair described by a pure Bell state, the entanglement per
  element constrained by this SSR goes to zero for a large number of
  elements.
\end{abstract}
\maketitle

Entanglement lies at the heart of quantum information processing
(QIP)~\cite{Nie00}, and quantifying entanglement as a physical
resource is a primary goal of this field~\cite{Hor00}.  Recently, it
has been shown that the existence of superselection rules
(SSRs)~\cite{Wic52} requires us to reassess traditional entanglement
measures~\cite{Wis02} and the allowed bipartite
operations~\cite{Ver03}: the SSRs enforce additional restrictions on
what Alice and Bob can accomplish using only local operations and
classical communication (LOCC).

In this Letter, we quantify entanglement constrained by a generalized
SSR and show that this entanglement is typically less than the amount
given by any standard measure.  To accomplish this task, we first
introduce the concept of a generalized SSR as a rule associated with
some group of physical transformations of a system.  The rule is
defined operationally: it restricts the allowed operations on the
system to those that are covariant with respect to that group.  Our
definition encompasses traditional SSRs such as charge and particle
number as well as effective SSRs for quantities such as angular
momentum or photon number (which may arise due to practical
restrictions on operations, the lack of an appropriate shared
reference frame, or through interaction with an environment).  Our
measure of entanglement constrained by SSRs is also operational in
that it describes the accessible entanglement that Alice and Bob can
distill into standard quantum registers through allowed LOCC.

As an explicit example of entanglement constrained by a SSR, we
describe ensemble QIP where access to individual elements of the
ensemble is not possible.  The relevant SSR here restricts the allowed
operations to be ``collective'' in that they act identically on all
elements of the ensemble.  We find that our operational measure of
entanglement constrained by this SSR can be hugely smaller than that
found from standard entanglement measures.  In particular, we prove
the remarkable result that, even if each element of the ensemble
consists of two qubits described by a \emph{pure} Bell state, the
entanglement per element constrained by this SSR is zero in the limit
of a large number of elements.  We discuss how this result places a
powerful constraint on QIP in liquid-state NMR~\cite{Cor97} and
spin-squeezing experiments~\cite{Jul01}.

We begin by providing an operational definition of a SSR, and show
that this definition is compatible with colloquial uses.  \emph{A SSR
  is a restriction on the allowed local operations on a system, and is
  associated with a group of physical transformations.}  This
restriction could be imposed by properties of the underlying theory
(e.g., a SSR for charge required in a Lorentz-invariant quantum field
theory~\cite{Str74}), but we also consider SSRs that arise due to
practical restrictions.  Consider a local quantum system with Hilbert
space $\mathbb{H}$.  The set of physical operations on this quantum
system is given by the semigroup of completely positive (CP)
trace-preserving maps $\{\mathcal{E}\}_{\rm CP}$.  These CP maps
describe not only unitary (closed) operations but also open processes
such as state preparation, dissipation and measurement.  Let $G$ be a
group of physical transformations acting on $\mathbb{H}$ through a
unitary representation $T$.  We define an operation $\mathcal{O} \in
\{\mathcal{E}\}_{\rm CP}$ to be \emph{$G$-covariant} if
\begin{equation}
  \label{eq:DefSSR}
  \mathcal{O}[T(g) \rho T^\dag(g)] = T(g)\mathcal{O}[\rho]T^\dag(g)
  \, , 
\end{equation}
for all group elements $g \in G$ and all density operators $\rho$.  We
then define the \emph{SSR associated with $G$}, or $G$-SSR, to be to
be a restriction on the allowed operations on the system to those CP
maps $\{ \mathcal{O} \}_{G\textrm{-SSR}} \subset \{\mathcal{E} \}_{\rm
  CP}$ that are $G$-covariant. The following examples reveal how this
definition is compatible with some traditional SSRs:

\emph{Example 1: Charge.}  Let $G$ be a one-dimensional Lie group U(1)
generated by a Hermitian operator $Q$; i.e., $T(\xi) =
\exp(\text{i}\xi Q)$.  If $Q$ is a local charge operator then this
U(1)-SSR is usually referred to as a SSR for charge. Similar SSRs can
be developed for particle number.  When such a SSR applies, one
cannot, for instance, {\em locally create} superpositions of charge
eigenstates because the required operations are not $G$-covariant.
Note that this SSR does not forbid the creation of superpositions
where, for example, one charge can be found at two different
locations, as in the twin-slit experiment for electrons.

\emph{Example 2: Angular momentum.}  Let $G=$ SO(3) be the rotation
group generated by the total angular momentum operators $\{ L_{x,y,z}
\}$.  The associated SSR ensures that all allowed operations are
rotationally invariant.  This SSR may apply, for instance, when there
is no reference frame for orientation and thus all observables commute
with the total angular momentum.  A reference frame would establish
operators that specify a direction; such operators do not commute with
total angular momentum and thus violate the SSR.  In this example, the
SSR is a practical rather than fundamental consideration: the lack of
a reference frame leads to a SSR.

\emph{Example 3: Environmentally-induced SSR.}  Let $\hat{H}_{\rm
  int}$ be a coupling Hamiltonian between the system and an
environment, and let $G=$ U(1) be the group generated by this
Hamiltonian.  Einselection~\cite{Zur03}, which is often expressed as
the condition that the only allowed states of the system are those
that commute with this Hamiltonian, has the form of a U(1)-SSR.

It should be noted that the existence of a SSR is not equivalent to a
conservation law; in fact, the only interesting SSRs are those that
apply to \emph{non-conserved} quantities~\cite{Wis03}.  Also note that
a SSR does not restrict the allowed states of the system.  However,
the restrictions imposed on the allowed operations by the $G$-SSR mean
that a state $\rho$ is indistinguishable from the states $T(g)\rho
T^\dag(g)$ for all $g \in G$.  Because of this indistinguishability,
it is operationally appropriate to describe $\rho$ by the state
\begin{equation}
  \label{eq:AveragedState}
  \mathcal{G}[\rho] \equiv \begin{cases} (\text{dim}\ G)^{-1}
  \sum_{g \in G}
  T(g) \rho T^\dag(g) \, ,  &\text{finite groups} \\ \int_G
  \text{d}v(g)\, T(g) \rho T^\dag(g) \, ,  &\text{Lie
  groups}\end{cases}  
\end{equation}
where $\text{d}v$ is the group-invariant (Haar) measure~\cite{Ful91}.
The state ${\cal G}[\rho]$ is invariant under the action of $G$,
\begin{equation}
  \label{eq:InvariantState}
  T(g) \mathcal{G}[\rho] T^\dag(g) = \mathcal{G}[\rho] \, , \quad
  \forall\ g \in G \, ,
\end{equation}
so we call this state the \emph{$G$-invariant state}.

Consider a SSR for charge as an example.  States that are
superpositions of charge eigenstates are not \emph{a priori}
prohibited.  For a state that is a superposition of charge eigenstates
$|\psi\rangle = \alpha |q_1\rangle + \beta |q_2\rangle$, with
$Q|q_i\rangle = q_i|q_i\rangle$, the effect of the superoperator
$\mathcal{G}$ on this state is
\begin{equation}
  \label{eq:MixChargeEigenstates}
  \mathcal{G}\bigl[|\psi\rangle\langle\psi|\bigr] 
  = |\alpha|^2 |q_1\rangle \langle q_1| + |\beta|^2|q_2\rangle\langle
  q_2 | \, .
\end{equation}
Thus, in the presence of the SSR, a superposition of charge
eigenstates is operationally equivalent to a \emph{mixture} of charge
eigenstates.  The effect of $\mathcal{G}$ is to project onto
eigenspaces of the group generator $Q$.  SSRs associated with
one-dimensional Lie groups are often defined this way ({\em
  cf}.~\cite{Ver03}).  However, for general SSRs (including the
example associated with the rotation group), there is not necessarily
an equivalent expression.
  
We now consider imposing a SSR in a bipartite setting; that is, both
parties (Alice and Bob) are restricted to local operations obeying
Eq.~(\ref{eq:DefSSR}).  Consider a bipartite state $\rho^{ab}$ shared
by Alice and Bob.  This state may have been prepared by a third party
under conditions where no SSR applies.  The $G$-invariant state
constrained by these local SSRs is $\mathcal{G}[\rho^{ab}] =
\mathcal{G}^a \otimes \mathcal{G}^b [\rho^{ab}]$.  To quantify the
entanglement of this state we assume that, in addition to this
bipartite system, Alice and Bob each possess a quantum register with
Hilbert space dimension equal to or greater than that of their
respective systems.  These registers are initially in the pure product
state $\varrho^{ab}_0$ and are not subject to any SSR.  (For example,
these registers could be standard qubits over which Alice and Bob have
complete control.)  We quantify the entanglement
$E_{G\textrm{-SSR}}(\rho^{ab})$ constrained by the $G$-SSR as the
maximum amount of entanglement that Alice and Bob can produce between
their registers by LOCC~\cite{Wis02}.  The latter can be quantified by
an appropriate standard measure $E$, e.g., the entanglement of
distillation~\cite{Nie00}.  The following theorem quantifies the
entanglement constrained by an arbitrary SSR for pure or mixed states,
generalizing the result of~\cite{Wis02}:

\textbf{Theorem:} The entanglement $E_{G\textrm{-SSR}}(\rho^{ab})$
that Alice and Bob can produce between their registers from the state
$\rho^{ab}$ by LOCC constrained by a $G$-SSR is given by the
entanglement $E(\mathcal{G}[\rho^{ab}])$ that they can produce from
the state $\mathcal{G}[\rho^{ab}]$ by {\em unconstrained} LOCC, where
$E$ is a standard measure of entanglement.

\textbf{Proof:} First, note that any CP map can be composed with
$\mathcal{G}$ to yield a $G$-invariant operation, i.e.,
\begin{equation}
  \label{eq:OasGE}
  \mathcal{G}\circ \mathcal{E}\circ
  \mathcal{G} \in \{ \mathcal{O} \}_{G\textrm{-SSR}} \quad \forall\
  \mathcal{E} \in \{ \mathcal{E}\}_{\rm CP} \, , 
\end{equation}
which follows from the definitions~(\ref{eq:DefSSR})
and~(\ref{eq:AveragedState}).  Let $\mathcal{O}$ be a $G$-invariant
operation in LOCC acting on the initial state $\rho^{ab} \otimes
\varrho^{ab}_0$.  The final state of the registers is given by
$\varrho^{ab}_{1} = {\rm Tr}_{\rm sys}\bigl( \mathcal{O} [\rho^{ab}
\otimes \varrho^{ab}_0] \bigr)$, where the trace is over the shared
system.  The maximum entanglement produced between the registers is
given by maximizing $E(\varrho^{ab}_{1})$ over all LOCC obeying the
$G$-SSR.  Thus, using~(\ref{eq:OasGE}),
\begin{align}
  \label{MaxEntTrans}
  E_{G\textrm{-SSR}}(\rho^{ab}) &= \max_{\mathcal{O}}\ E\bigl({\rm
    Tr}_{\rm sys}\bigl(
  \mathcal{O}[\rho^{ab} \otimes \varrho^{ab}_0] \bigr) \bigr) \nonumber \\
  &= \max_{\mathcal{O}}\ E\bigl({\rm Tr}_{\rm sys} \bigl(
  (\mathcal{G}\circ \mathcal{O}\circ\mathcal{G}) [\rho^{ab} \otimes
  \varrho^{ab}_0]
  \bigr) \bigr) \nonumber \\
  &= \max_{\mathcal{E}}\ E\bigl({\rm Tr}_{\rm sys} \bigl(
  (\mathcal{G}\circ \mathcal{E}\circ\mathcal{G}) [\rho^{ab} \otimes
  \varrho^{ab}_0]
  \bigr) \bigr) \nonumber \\
  &= \max_{\mathcal{E}}\ E\bigl({\rm Tr}_{\rm sys} \bigl( \mathcal{E}
  \bigl[\mathcal{G} [\rho^{ab}] \otimes \varrho^{ab}_0\bigr]
    \bigr)\bigr) \, , 
\end{align}
where the second line follows from the properties of trace and the
definition~(\ref{eq:AveragedState}), and the last line follows from
the properties of trace.  The latter maximization is over {\em all}
LOCC (not just operations in $\{ \mathcal{O} \}_{G\textrm{-SSR}}$),
and gives the entanglement $E(\mathcal{G}[\rho^{ab}])$ that Alice and
Bob can produce between their registers from the state
$\mathcal{G}[\rho^{ab}]$ by unconstrained LOCC. \hfill$\Box$\medskip

We now turn to a specific application of the above result that yields
a striking difference between the amount of entanglement with and
without the SSR.  Ensemble QIP describes $N$ identical copies of a
system of qubits, where $N$ is usually taken to be very large.  We
consider a situation where access to individual elements of the
ensemble is not possible, and thus only \emph{collective}
transformations and measurements (i.e., operations which affect each
element identically) are allowed.  In the following, we formulate this
restriction as a SSR, and show that it severely limits the
entanglement in the system.

Consider an ensemble consisting of $N$ copies of a single qubit. (For
convenience, we assume $N$ is even.)  The Hilbert space
$\mathbb{H}_2^{\otimes N}$ carries a collective tensor representation
$R$ of SU(2), by which a rotation $\Omega \in$ SU(2) acts identically
on each of the $N$ qubits.  The Hilbert space also carries a
representation $P$ of the symmetric group $S_N$, which is the group of
permutations of the $N$ qubits.  The action of these two groups
commute, and Schur-Weyl duality~\cite{Ful91} states that the Hilbert
space $\mathbb{H}_2^{\otimes N}$ carries a multiplicity-free direct
sum of SU(2)$\times S_N$ irreducible representations (irreps), each of
which can be labelled by the SU(2) total angular momentum quantum
number $j$.  Each of these irreps can be factored into a tensor
product $\mathbb{H}_{jR} \otimes \mathbb{H}_{jP}$, such that SU(2)
acts irreducibly on $\mathbb{H}_{jR}$ and trivially on
$\mathbb{H}_{jP}$, and $S_N$ acts irreducibly on $\mathbb{H}_{jP}$ and
trivially on $\mathbb{H}_{jR}$.  Thus,
\begin{equation}
  \label{eq:DirectSumOfProducts}
  \mathbb{H}_2^{\otimes N} = \bigoplus_{j=0}^{N/2} \mathbb{H}_{jR}
  \otimes \mathbb{H}_{jP} \, .
\end{equation}
The dimension of $\mathbb{H}_{jR}$ is $2j+1$, and that of
$\mathbb{H}_{jP}$ is~\cite{Bar03}
\begin{equation}
   c_j^{(N)}=\binom{N}{N/2-j}\frac{2j+1}{N/2+j+1}.
\end{equation}
Consider a basis $|j,m\rangle_R \otimes |j,r\rangle_P$ for
$\mathbb{H}_{jR} \otimes \mathbb{H}_{jP}$, with $\{ |j,m\rangle_R,
m=-j,\ldots,j\}$ the standard angular momentum basis for
$\mathbb{H}_{jR}$ and $\{ |j,r\rangle_P, r=1,\ldots, c_j^{(N)} \}$ a
basis for $\mathbb{H}_{jP}$.  The group SU(2)$\times S_N$ acts on this
basis as $R(\Omega)|j,m\rangle_R \otimes P(p)|j,r\rangle_P$ for
$\Omega \in$ SU(2) and $p \in S_N$.

In ensemble QIP without individual addressability, the only allowed
operations $\mathcal{O}$ are those that are invariant under
permutations of elements and thus must satisfy
\begin{equation}
  \label{eq:NMRSSR}
  \mathcal{O}[P(p) \rho P^\dag(p)] = P(p)\mathcal{O}[\rho]P^\dag(p)
  \, , 
\end{equation}
for all $p \in S_N$.  (Note that these allowed operations include all
transformations generated by Hamiltonians in the enveloping algebra of
su(2), i.e., which are polynomials in $J_x$, $J_y$ and $J_z$. In
liquid-NMR QIP the operations are more restricted because there are no
controllable inter-molecular interactions.  We will not consider that
additional restriction here.)  Thus, ensemble QIP with this
restriction must respect the SSR associated with the symmetric group
$S_N$.  Unlike previous examples involving Lie groups, this SSR is
associated with a finite group.  We define the superoperator
\begin{equation}
  \label{eq:MixAllPermutations}
  \mathcal{P}[\rho] = \frac{1}{N!} \sum_{p \in S_N} P(p) \rho
  P^\dag(p) \, ,
\end{equation}
which can be extended to act on states of $n$ qubits on $N$ elements.
The action of $\mathcal{P}$ is best seen in the decomposition of
Eq.~(\ref{eq:DirectSumOfProducts}): it completely mixes states in
$\mathbb{H}_{jP}$ while leaving states in $\mathbb{H}_{jR}$ invariant.
The spaces $\mathbb{H}_{jR}$ are called noiseless subsystems
(NSs)~\cite{Kni00}, and are free from the decohering effect of
$\mathcal{P}$.  These NSs are dual to the collective NSs
$\mathbb{H}_{jP}$, which have been explored in the context of quantum
computation~\cite{Kni00,Zan97} and quantum communication without shared
orientation reference frames~\cite{Bar03} and which are free from
collective SU(2) decoherence.

We now quantify bipartite entanglement in ensemble QIP constrained by
the $S_N$-SSR; specifically, we show that the standard (unconstrained)
measure of entanglement can grossly overestimate the amount of
entanglement accessible to Alice and Bob.  Consider the following
example, where each of $N$ elements possesses two qubits, $a$ and $b$.
These qubits are separated such that all the qubits of type $a$ are
given to Alice and $b$ to Bob.  If the $S_N$-SSR is enforced, Alice
and Bob are both restricted to local $S_N$-covariant operations, and
any state $\rho^{ab}$ is indistinguishable from the state
$\mathcal{P}[\rho^{ab}] = \mathcal{P}^a \otimes
\mathcal{P}^b[\rho^{ab}]$.

If the two qubits on every element are described by the Bell state
$|\Phi^+\rangle = \frac{1}{\sqrt{2}}\bigl(|uu\rangle^{ab} +
|dd\rangle^{ab} \bigr)$, the state of the total ensemble is
$|\Phi^{(N)}\rangle \equiv |\Phi^+\rangle^{\otimes N}$.  A naive
quantification of entanglement of this pure state gives $N$ ebits.
However the constrained entanglement is much less.  Expressing this
state in terms of the decomposition of
Eq.~(\ref{eq:DirectSumOfProducts}) yields
\begin{align}
  \label{eq:NMaxEntangled2}
  |\Phi^{(N)}\rangle &= \frac{1}{\sqrt{2^N}} \sum_{j=0}^{N/2}
   \sum_{m=-j}^j \sum_r |j,m\rangle_{R}^a |j,r \rangle_{P}^a
   |j,m\rangle_{R}^b |j,r \rangle_{P}^b \nonumber \\
  &= \sum_{j=0}^{N/2} \sqrt{\frac{(2j+1)c_j^{(N)}}{2^N}}
   |\phi_j\rangle^{ab} |\chi_j\rangle^{ab} \, , 
\end{align}
where
\begin{align}
  \label{eq:MaxEntj}
  |\phi_j\rangle^{ab} &= \frac{1}{\sqrt{2j+1}} \sum_{m=-j}^j
  |j,m\rangle_{R}^a  |j,m\rangle_{R}^b \, , \\
  |\chi_j\rangle^{ab} &= \frac{1}{\sqrt{c_j^{(N)}}} \sum_{r} 
  |j,r\rangle_{P}^a  |j,r\rangle_{P}^b \, ,
\end{align}
are (normalised) maximally entangled states in $\mathbb{H}_{jR}^a
\otimes \mathbb{H}_{jR}^b$ and $\mathbb{H}_{jP}^a \otimes
\mathbb{H}_{jP}^b$, respectively.

The action of $\mathcal{P}$ on the state (\ref{eq:NMaxEntangled2}) is
\begin{equation}
  \mathcal{P}\bigl[|\Phi^{(N)}\rangle
  \langle\Phi^{(N)}| \bigr] 
  = \sum_{j=0}^{N/2} \frac{(2j+1)c_j^{(N)}}{2^N}
  |\phi_j\rangle^{ab}\langle\phi_j| 
  \otimes \sigma_{j}^{ab} \, , 
\end{equation}
where $\sigma_{j}^{ab}$ is the (normalised) completely mixed state on
$\mathbb{H}_{jP}^a \otimes \mathbb{H}_{jP}^b$.  The resulting state is
an incoherent sum of maximally-entangled states on each irrep
$\mathbb{H}_{jR}^a \otimes \mathbb{H}_{jR}^b$.  The entanglement of
this state can be easily calculated because both Alice and Bob can
\emph{locally} perform a measurement of total $J^2$, which determines
$j$ and yields a pure state in $\mathbb{H}_{jR}^a \otimes
\mathbb{H}_{jR}^b$.  Thus, the entanglement of this state constrained
by the $S_N$-SSR is
\begin{equation}
  \label{eq:EntOfNBellStates}
  E_{S_N\textrm{-SSR}} \bigl(|\Phi^{(N)}\rangle\langle\Phi^{(N)}|\bigr) = %
  \sum_{j=0}^{N/2} \frac{(2j+1) c_j^{(N)}}{2^N} \log_2 (2j+1) \, , 
\end{equation}
which, for large $N$, behaves as $\frac{1}{2}\log_2 N$.  Thus,
although the state $|\Phi^{(N)}\rangle$ possesses $N$ ebits of
unconstrained entanglement, its entanglement constrained by the SSR is
only $\frac{1}{2}\log_2 N$ ebits (asymptotically).  This result is
remarkable: for a pure state of the ensemble, each element consisting
of two qubits in a Bell state, the constrained entanglement per
element rapidly approaches zero for large $N$.

Note that the state $|\Phi^{(N)}\rangle$, describing each element as a
Bell state, is \emph{not} a maximally entangled state under the
$S_N$-SSR constraint; we now identify such a state.  Observing the
form of the decohering mechanism $\mathcal{P}$, this state is clearly
a pure maximally entangled state in the NS $\mathbb{H}_{jR}^a \otimes
\mathbb{H}_{jR}^b$ with the largest dimension, given by $j_0 =N/2$.
The entanglement per element of this state is $N^{-1} \log_2 (N+1)$,
which approaches zero for large $N$.  As this state is the maximally
entangled state under the SSR constraint, we have proved that the
maximum entanglement per element in the large $N$ limit is zero.

A remarkable duality is evident between the SSR associated with the
symmetric group and that associated with the rotation group when one
considers the maximum entanglement these SSRs allow. The latter
describes a situation where Alice and Bob do not share an orientation
reference frame for their qubits~\cite{Bar03}.  In the Hilbert space
decomposition of Eq.~(\ref{eq:DirectSumOfProducts}), the SSR for the
rotation group is described by a decohering superoperator on the SU(2)
irreps, as opposed to the $S_N$ irreps described above.  As proven
in~\cite{Bar03}, the maximum entanglement between $N$ pairs of qubits
constrained by the SU(2)-SSR behaves asymptotically as $N - \log_2 N$;
in this letter, we proved that the maximum entanglement between $N$
pairs of qubits constrained by the $S_N$-SSR behaves asymptotically as
$\log_2 N$.  Note that these two values asymptotically sum to $N$,
which is the maximum unconstrained entanglement for $N$ qubit pairs.

In summary, we have defined generalized SSRs, and quantified the
constrained entanglement of a bipartite state as the amount of
entanglement that can be distilled into quantum registers using only
LOCC that obey the appropriate SSR.  Our example of ensemble QIP
reveals that systems with apparently large amounts of entanglement can
in fact possess very little under the appropriate SSR constraint. We
note that this result for ensemble QIP applies to liquid-state
NMR~\cite{Cor97}, where qubits are realized as nuclear spins on a
molecule and a sample generally contains $N\sim 10^{20}$
molecules~\footnote{However, NMR QIP possesses additional constraints
  due to impure initial conditions, restrictions on the allowed
  operations, and inefficient read-out.}.  Our operational definition
of entanglement constrained by a $S_N$-SSR is also applicable to
spin-squeezed atomic gases~\cite{Jul01}; because the NS corresponding
to $j_0 =N/2$ is used in such experiments, our result shows that the
present measures of entanglement used for this system~\cite{Sto03}
\emph{are} appropriate.

Another question related to our operational definition of entanglement
is: What constraints does a SSR impose on the entangled states that
can be created (formed) from a set amount of entanglement in the
quantum registers?  As pointed out in~\cite{Ver03}, it is not even
possible to create certain \emph{separable} states in the presence of
a SSR.  It is clear that SSRs place severe contraints on QIP, and our
operational definitions of SSRs and entanglement constrained by them
provide a new understanding and a valuable tool to quantum information
science.

\begin{acknowledgments}
  We thank E.~Knill, R.~W.~Spekkens, F.~Verstraete and P.~Zanardi for
  helpful comments on our manuscript, and thank M.~A.~Nielsen,
  T.~Rudolph, B.~C.~Sanders, and R.~W.~Spekkens for valuable prior
  discussions.
\end{acknowledgments}

\end{document}